\documentclass[twocolumn]{aastex631}
\usepackage{multirow}
\usepackage{xspace}

\usepackage{amsmath}
\usepackage{bm}
\usepackage{comment}

\newcommand{\Msun}{\,{\rm M_\odot}}

\newcommand{\Mblack}{M_\bullet}
\newcommand{\Mstar}{M_\star}
\newcommand{\Ha}{\rm H\alpha}
\newcommand{\mmstar}{$\Mblack-\Mstar$\xspace}

\begin{document}

\title{JWST CEERS \& JADES Active Galaxies at $z = 4-7$ Violate the Local \mmstar Relation at $>3\sigma$: \\ Implications for Low-Mass Black Holes and Seeding Models}

\correspondingauthor{Fabio Pacucci}
\email{fabio.pacucci@cfa.harvard.edu}

\author[0000-0001-9879-7780]{Fabio Pacucci}
\affiliation{Center for Astrophysics $\vert$ Harvard \& Smithsonian, Cambridge, MA 02138, USA}
\affiliation{Black Hole Initiative, Harvard University, Cambridge, MA 02138, USA}

\author[0000-0002-7524-5219]{Bao Nguyen}
\affiliation{Steward Observatory, University of Arizona, 933 N. Cherry Ave., Tucson, AZ 85721, USA}
\affiliation{Department of Physics, University of Arizona, 1118 East Fourth Street, Tucson, AZ 85721, USA}

\author[0000-0002-6719-380X]{Stefano Carniani}
\affiliation{Scuola Normale Superiore, Piazza dei Cavalieri 7, I-56126 Pisa, Italy}

\author[0000-0002-4985-3819]{Roberto Maiolino}
\affiliation{Kavli Institute for Cosmology, University of Cambridge, Madingley Road, Cambridge, CB3 0HA, UK}
\affiliation{Cavendish Laboratory, University of Cambridge, 19 JJ Thomson Avenue, Cambridge CB3 0HE, UK}
\affiliation{Department of Physics and Astronomy, University College London, Gower Street, London WC1E 6BT, UK}

\author[0000-0003-3310-0131]{Xiaohui Fan}
\affiliation{Steward Observatory, University of Arizona, 933 N. Cherry Ave., Tucson, AZ 85721, USA}



\begin{abstract}
JWST is revolutionizing our understanding of the high-$z$ Universe by expanding the black hole horizon, looking farther and to smaller masses, and revealing the stellar light of their hosts. By examining JWST galaxies at $z=4-7$ that host $\Ha$-detected black holes, we investigate (i) the high-$z$ \mmstar relation and (ii) the black hole mass distribution, especially in its low-mass range ($\Mblack \lesssim 10^{6.5} \Msun$).
With a detailed statistical analysis, our findings conclusively reveal a high-$z$ \mmstar relation that deviates at $>3\sigma$ confidence level from the local relation. The high-$z$ relation is: $\log(\Mblack/\Msun) = -2.43^{+0.83}_{-0.83} + 1.06^{+0.09}_{-0.09} \log(\Mstar/\Msun)$. Black holes are overmassive by $\sim 10-100\times$ compared to their low-$z$ counterparts in galactic hosts of the same stellar mass. This fact is not due to a selection effect in surveys. 
Moreover, our analysis predicts the possibility of detecting in high-$z$ JWST surveys $5-15\times$ more black holes with $\Mblack \lesssim 10^{6.5} \Msun$, and $10-30\times$ more with $\Mblack \lesssim 10^{8.5} \Msun$, compared to local relation's predictions. The lighter black holes preferentially occupy galaxies with a stellar mass of $\sim 10^{7.5}-10^8 \Msun$. We have yet to detect these sources because (i) they may be inactive (duty cycles $1\%-10\%$), (ii) the host overshines the AGN, or (iii) the AGN is obscured and not immediately recognizable by line diagnostics. A search of low-mass black holes in existing JWST surveys will further test the \mmstar relation. Current JWST fields represent a treasure trove of black hole systems at $z = 4-7$; their detection will provide crucial insights into their early evolution and co-evolution with their galactic hosts.
\end{abstract}


\section{Introduction} 
\label{sec:intro}
The James Webb Space Telescope (JWST) was designed to be ``a giant leap forward in our quest to understand the Universe and our origins''. 
Faithful to its core mission, JWST shows us an unprecedented — and somewhat unexpected — view of the high-$z$ Universe, particularly concerning early galaxy and black hole formation.

Before the advent of JWST observations, the ``black hole horizon,'' defined as the farthest redshift where a black hole can be detected, stood at $z \sim 7-8$. The farthest supermassive black hole (SMBH), with mass $\Mblack = (1.6 \pm 0.4) \times 10^9 \Msun$, was detected at $z=7.642$ \citep{Wang_2021_quasar}. In its first year of operation, JWST immensely expanded the black hole horizon and significantly decreased the detected black hole mass. 

Currently, the farthest SMBH is observed spectroscopically in GN-z11, a galaxy at $z=10.6$, with a mass of $\sim 1.6 \times 10^6 \Msun$ \citep{Maiolino_2023}. This discovery came shortly after the $\sim 2 \sigma$ detection of another quasar at $z=8.679$, with a slightly heavier mass of $\sim 10^7 \Msun$ \citep{Larson_2023}.  

These exceptional discoveries come on top of more numerous detections in the redshift range $z = 4-7$ (i.e., where the $\Ha$ line is effective in measuring masses) that is filling the $10^6-10^8 \Msun$ black hole mass range \citep{Harikane_2023, Kocevski_2023, Onoue_2023, Carnall_2023, Ubler_2023}.
This redshift range is now filled with many galaxies hosting black holes, also of small mass, that allow some statistical inference. 

The JWST is also revolutionizing the field in another aspect: revealing the starlight emitted by the hosts of SMBHs. Although we detected more than $200$ quasars at $z>6$ \citep{Fan_2022_review}, these SMBHs are very massive, accrete at or near the Eddington luminosity, and typically outshine their host galaxy, with typical luminosities of $>10^{46} \, \mathrm{erg \, s^{-1}}$. Thus, detecting the host's starlight of SMBHs during the reionization epoch has been elusive. With the photon-collecting power of JWST, we can now detect SMBHs that are smaller in mass, farther out, and characterized by lower luminosities ($\sim 10^{44}-10^{45} \, \mathrm{erg \, s^{-1}}$). As they do not outshine their hosts, making estimates of their stellar mass is now achievable, with relative ease, in this redshift range \citep{Ding_2022}.

What is the astrophysical picture that these observations provide us? Expanding the black hole horizon in redshift and decreasing the detected black hole mass enlighten us regarding two crucial astrophysical topics:
\begin{enumerate}
    \item The formation of the first population of black holes, referred to as black hole seeds.
    \item The early co-evolution of black holes and their galactic hosts.
\end{enumerate}

Black hole seeds formed at $z \sim 20-30$, or $\sim 200$ Myr after the Big Bang, \citep{BL01}, when the first population of stars (Pop III) were born.
Seeds are typically categorized into light ($\Mblack \sim 10^2 \Msun$) and heavy ($\Mblack \sim 10^5 \Msun$) seeds (see, e.g., the reviews  \citealt{Woods_2019, Inayoshi_review_2019, Fan_2022_review}).
By detecting black holes at progressively higher redshift, we automatically shrink the time frame between detection and formation, thus constraining more tightly the properties of early seeds \citep{Pacucci_2022_search, Fragione_Pacucci_2023}.

The existence of a clear correlation, at least in the local Universe, between the mass of the central SMBH and some properties of its host indicates that these two crucial cosmic players co-evolved \citep{Magorrian_1998, Ferrarese_Merritt_2000, Gebhardt_2000, Kormendy_Ho_2013, Reines_Volonteri_2015, Shankar_2019}. These correlations (e.g., the \mmstar between the SMBH mass and the host's stellar mass) have been tested at low redshift and also scrutinized at higher redshift, although with more uncertain and scarce data (see, e.g., \citealt{Volonteri_Reines_2016} and more recently with JWST data, e.g., \citealt{Maiolino_2023, Larson_2023, Kocevski_2023, Stone_2023, Kokorev_2023, Yue_2023}). Crucially, \cite{Volonteri_2023} posed the timely question of whether $z > 9$ galaxies recently discovered by JWST host a central black hole and what their masses might be. 

At these redshifts, the validity of the \mmstar is uncertain, while other relations, e.g., the $\Mblack-\sigma$ with the central stellar velocity dispersion, and the $\Mblack-M_{\rm dyn}$ with the dynamical mass of the host, seem to hold \citep{Maiolino_2023}.
In this study, whenever we use the term ``overmassive'', we refer to black holes that are too massive compared to what their \textit{stellar mass} would dictate.

The aforementioned fundamental unknowns of the early cosmic epochs are connected. Several studies have shown that, in the high-$z$ Universe, offsets from local relations may favor specific flavors of black hole seeds (see, e.g., \citealt{Hirschmann_2010, Pacucci_MaxMass_2017, Visbal_Haiman_2018, Pacucci_2018, Nguyen_2019, Greene_2020_review, King_2021, Hu_2022, Koudmani_2022, Schneider_2023, Scoggins_2023, Scoggins_2023_2}). Hence, constraining the shape of the \mmstar relation at high-$z$, especially its low-mass end (i.e., with sources that are more common than quasars), will enlighten us on the seeding mechanisms during earlier epochs.

Using new JWST observations of galaxies hosting SMBHs at $z = 4-7$, we pose two fundamental questions: (i) What do current JWST observations imply regarding the high-$z$ scaling relation between black hole and stellar mass? (ii) How many SMBHs do we expect to observe with JWST in current and future surveys, especially in the lower end of the mass distribution? 

\section{Data}
\label{sec:data}

We examine the following data set of SMBHs and their host galaxies:
\begin{itemize}
    \item 8 sources reported by \cite{Harikane_2023}, obtained from the Early Release Observations \citep{Pontoppidan_2022}, which targeted the SMACS 0723 field, the Early Release Science observations of GLASS \citep{Treu_2022}, and the Cosmic Evolution Early Release Science Survey (CEERS, \citealt{Finkelstein_2023}). From this sample, we included only sources that are not gravitationally lensed because lensed objects follow different detection limits. Note that this set already includes the two sources reported by \cite{Kocevski_2023}, with CEERS identification numbers 2782 and 746.
    \item 12 sources reported by \cite{Maiolino_2023_new} from the JWST Advanced Deep Extragalactic Survey (JADES, \citealt{JADES_2023}). Three of these 12 sources are indicated as candidate dual AGN, and the mass of the secondary SMBH is also provided. As these secondary SMBHs are candidates, we decided not to include them in this analysis.
    \item 1 source reported by \cite{Ubler_2023} from the Galaxy Assembly with NIRSpec IFS survey (GA-NIFS).
\end{itemize}

We chose these sources because of two critical commonalities: they are all spectroscopically confirmed with NIRSpec, and their black hole masses are estimated with the $\Ha$ line \citep{Greene_2005}. Note that NIRSpec provides near-IR spectroscopy up to $5.3 \, \rm \mu m$; hence this method can be applied to estimate SMBH masses up to $z\sim 7$. All the sources investigated here are in the redshift range $4<z<7$ and are characterized by an estimate of the hosts' stellar mass, which is executed with a fitting of their UV to optical rest-frame spectral energy distributions.

As this work analyzes the possible bias introduced by observations, we exclude the two black holes at $z>8$ \citep{Maiolino_2023, Larson_2023} whose masses are estimated with other optical lines and require a different analysis.

\section{Methods}
\label{sec:methods}

In this Section, we describe the analytical and statistical tools used to analyze the data and infer the properties of the population of $z=4-7$ galactic systems.

\subsection{Local Scaling Relation}
\label{subsec:local_mmstar}
We consider the relation derived by \cite{Reines_Volonteri_2015} as our benchmark for the \mmstar correlation in the local Universe.
We adopted this relation because it was empirically estimated from a sample of bright AGN hosted in galaxies with stellar masses comparable with those of our sample and because it also uses the broad $\Ha$ line to estimate black hole masses between $10^6 \Msun$ and $10^8 \Msun$, which is compatible with the range investigated here. On the other hand, the standard relation \cite{Kormendy_Ho_2013} was determined from local, massive, and quiescent galaxies hosting non-accreting black holes at their centers.
The \cite{Reines_Volonteri_2015} relation is the following:
\begin{equation}
    \log \left( \frac{\Mblack}{\Msun} \right) = \alpha + \beta \log \left (\frac{\Mstar}{10^{11} \Msun} \right) \, ,
\end{equation}
where $\alpha = 7.45 \pm 0.08$ and $\beta = 1.05 \pm 0.11$.
This relation indicates that, at $z \sim 0$, the black hole mass is typically $0.1\%$ of the stellar mass contained in the bulge of its host.

Recent studies (e.g., \citealt{Kocevski_2023, Harikane_2023, Ubler_2023, Maiolino_2023, Maiolino_2023_new}) have also used the relation in \cite{Reines_Volonteri_2015} to compare the behavior of their high-$z$ systems with $z\sim 0$ ones. They find indications that the black holes grow faster than the stellar content of their host galaxy at $z > 4$. Interestingly, \cite{Bogdan_2023} reported the discovery of a gravitationally-lensed quasar at $z=10.3$, with an estimated black hole mass comparable to its stellar mass, and \cite{Natarajan_2023} suggested its origin as a heavy black hole seed. However, the detection of overmassive black holes is not new, as it was already suggested by sporadic observations at $z\sim 6$ in the pre-JWST era (see, e.g., \citealt{Wang_2010}), and, recently, even more locally (see, e.g., \citealt{Mezcua_2023}, up to $z=0.9$).

\subsection{JWST Sensitivity in $\Ha$ and Mass Limits}
\label{subsec:JWST_sensitivity}
All SMBHs in our sample were detected, and their mass estimated, using the $\Ha$ line emission (see Sec. \ref{sec:data}). To compare our analysis with the local \mmstar relation derived by \cite{Reines_Volonteri_2015}, we use the formula derived by \cite{Reines_2013} to estimate the black hole mass:
\begin{equation}
\begin{split}
    \log \left( \frac{\Mblack}{\Msun} \right) &= 6.60 + 0.47 \log \left( \frac{L_{\Ha}}{10^{42} \, \rm erg \, s^{-1}}\right) + \\ 
    &+ 2.06 \log \left( \frac{{\rm FWHM}_{\Ha} }{10^{3} \, \rm km \, s^{-1}}\right) \, .
\label{mass_Ha}
\end{split}
\end{equation}
We recalculate the black hole masses in our sample that are not estimated by Eq. \ref{mass_Ha} according to this relation for consistency. This procedure was also followed by \cite{Maiolino_2023_new}. Note that $L_{\Ha}$ and the full-width at half maximum (FWHM) refer to the component associated with the broad line region (BLR) of the AGN. 

To determine the sensitivity limit of JWST, we require NIRSpec to perform a $3\sigma$ detection, or better, to positively identify a BLR and, thus, a black hole at the center of its host.
For a medium JWST program (e.g., CEERS and JADES), the sensitivity for a resolution element of the spectrum is $\sim 10^{-19} \, \rm erg \, s^{-1} \, cm^{-2}$. Considering a resolution element of $\sim 300 \, \rm km \, s^{-1}$, and an FWHM of $1000 \, \rm km\, s^{-1}$ (the minimum FWHM for a BLR identified in \citealt{Harikane_2023}), the sensitivity on an integrated line is $\approx 1.8\times 10^{-19} \, \rm erg \, s^{-1} \, cm^{-2}$, or $\approx 5.5 \times 10^{-19} \, \rm erg \, s^{-1} \, cm^{-2}$ for a $3\sigma$ detection.

From this limiting flux, we compute the $L_{\Ha}$ luminosity (at different redshifts) and thus estimate the minimum black hole mass for which NIRSpec can execute a $3\sigma$ detection. We adopt the standard $\Lambda$CDM cosmology from \cite{Planck_2018} to find the luminosity distance $d_L(z)$. 
Additionally, NIRSpec must discern the $\Ha$ FWHM from the host's emission to accurately estimate the black hole mass. Considering typical radii of $z=4-7$ galaxies discovered by JWST and estimating their velocity dispersion from local galaxies, we calculated that the host's dispersion does not affect the $\Ha$ sensitivity thus far calculated for galaxies with a stellar mass $\Mstar < 10^{11} \Msun$. Hence, our sensitivity estimates based on the $\Ha$ luminosity are correct in the mass range we consider.

\begin{figure}%
    \centering
\includegraphics[angle=0,width=0.5\textwidth]{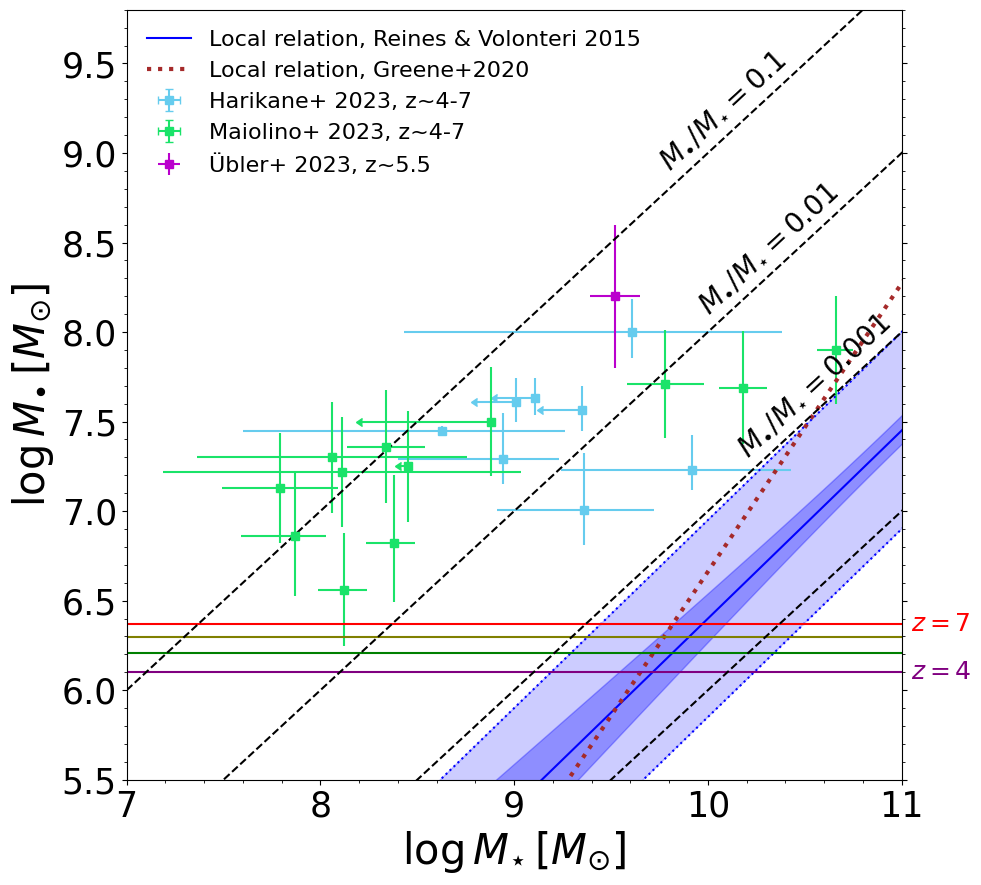}
    \caption{Overview of the data set used and the local \mmstar relation from \cite{Reines_Volonteri_2015}. The dark-shaded region denotes the $1\sigma$ uncertainty from the linear fit, and the light-shaded region denotes the 0.55 dex root-mean-square deviation of the local observations from the local relation. The black hole masses are estimated from the $\Ha$ virial relation from \cite{Reines_2013} with the parameters chosen by \cite{Reines_Volonteri_2015}. At four different redshifts ($z=4, 5, 6, 7$), the horizontal lines show the limiting sensitivities for a $3\sigma$ detection of the $\Ha$ line with NIRSpec (see Sec. \ref{subsec:JWST_sensitivity} for a detailed description). Error bars indicate the $1\sigma$ uncertainty in the stellar and black hole masses. As a comparison, the dotted, brown line displays the \mmstar relation by \cite{Greene_2020_review}.}
    \label{fig:limits}%
\end{figure}

Figure \ref{fig:limits} shows the data set used, along with the limiting sensitivities for NIRSpec, calculated at $z = 4, 5, 6, 7$. All data points are above the limiting sensitivities at their respective detection redshift. The figure also highlights the difference between the \mmstar relation by \cite{Reines_Volonteri_2015} and by \cite{Greene_2020_review}. The latter is even more in tension with high-$z$ JWST data than the former.

\subsection{High-$z$ Galaxy Stellar Mass Function}
\label{subsec:GMF}

Galaxies are not equally distributed in stellar mass. To estimate the likelihood of JWST detecting hosts at different stellar masses, we use the galaxy stellar mass function (SMF) in the range $z=4-8$ by \cite{Song_2016}:
\begin{equation}
    \phi(\Mstar, z) = \frac{\phi^{\ast}}{M^{\ast}} \left(\frac{\Mstar}{M^{\ast}}\right)^{\alpha} \exp{\left(-\frac{\Mstar}{M^{\ast}}\right)},
\label{smf}
\end{equation}
where the parameters $M^{\ast}$, $\phi^{\ast}$, and $\alpha$ are calibrated for the different redshifts considered. This SMF is derived from the ultraviolet luminosity function based on observations in the CANDELS/GOODS and Hubble Ultra Deep fields \citep{Song_2016}. The SMF is expressed in units of Mpc$^{-3}$ dex$^{-1}$; hence, to convert into a surface density (arcmin$^{-2}$ dex$^{-1}$) for different redshift bins of $\Delta z = 1$, we multiply the SMF with the volume enclosing a solid angle $\Omega_{sq'}$ within the redshift bin:
\begin{equation}
    V(z) = [V_{C}(z+0.5) - V_{C}(z-0.5)] \, \frac{\Omega_{sq'}}{4\pi},
\end{equation}
where $V_{C}(z)$ indicates the comoving volume at redshift $z$, assuming the \cite{Planck_2018} cosmology.

\subsection{Likelihood Function and Estimate of the High-z \mmstar Relation}
\label{subsec:likelihood}
To model the likelihood of the parameters of the high-$z$ \mmstar relation, given JWST data, we consider the following factors:

\begin{itemize}
    \item A two-dimensional Gaussian uncertainty distribution along the \mmstar relation, derived from the correlated stellar and black hole mass measurement uncertainties of our sample (Sec. \ref{sec:data}).
    \item An intrinsic Gaussian scatter around the \mmstar relation, with orthogonal variance $\nu$ as a free parameter. This scatter provides flexibility in fitting a linear relationship, considering the large uncertainties characterizing high-$z$ sources.
    \item The SMF (Eq. \ref{smf}) at $z = 5$, the median redshift in our data set, which increases the statistical weight to sources at lower stellar masses, where the population is expected to be more numerous.
    \item As we map the black hole mass function to the SMF, we are intrinsically assuming an occupation fraction of unity: all galaxies in our redshift range host an active black hole. Although much uncertainty remains, typical values for the AGN fraction in early galaxies range from $1\%$ to $10\%$ \citep{Harikane_2023_AGN_frac, Matthee_2023, Maiolino_2023_new}.
    \item The black hole mass limit based on the $\Ha$ line (see Sec. \ref{subsec:JWST_sensitivity}) at $z = 5$, which takes the bias from a flux-limited survey (the Lauer bias, see \citealt{Lauer_2007}) into account and increases the chance of detecting overmassive black holes.
    \item The single JWST observations are assumed to be statistically independent.
\end{itemize}

Although the mass uncertainties are reported to be non-Gaussian (i.e., with different upper and lower error bars), we model them as Gaussian for simplicity, using the greater error bars as a conservative estimate of standard deviations. The two-dimensional uncertainty distribution and the intrinsic scatter around the \mmstar relation are convoluted and projected orthogonally onto the relation, resulting in the following likelihood function for a sample of $N$ observations \citep{Hogg_2010}:

\begin{equation}
\begin{split}
    \ln {\cal L} &= \sum_{i = 1}^{N} \biggr[- \frac{\Delta_{i}^2}{2(\Sigma_{i}^2 + \nu)} - \frac{1}{2} \ln (\Sigma_{i}^2 + \nu) \, + \\ 
    &+ \ln \phi_i(x, z = 5) + \ln \eta_i \biggr] \, ,
\label{ln_L}
\end{split}
\end{equation}
where
\begin{subequations}
\begin{align}
    \Delta y_i &= y_i - m x_i - b \\
    \Delta x_i &= \Delta y_i \sin \theta \cos \theta \\
    t_i &= [y_{\text{lim}} - m(x_i + \Delta x_i) - b] \sec \theta \\
    \Delta_{i} &= \Delta y_i \cos \theta \\
    \Sigma_{i} &= (\sigma_x^2 \sin^2 \theta + \sigma_y^2 \cos^2 \theta - 2 \sigma_{xy} \sin \theta \cos \theta)^{1/2} \\
    \eta_i &= [1 - F(t_i)]^{-1}
\end{align}
\end{subequations}

Here, we provide only a short description of the terms involved; the interested reader is referred to \cite{Hogg_2010} for a detailed explanation.
In the formulas above, $i$ denotes each observation in the sample. We define $x = \log \Mstar$ and $y = \log \Mblack$ in units of solar mass. $m$ and $b$ denote the slope and y-intercept of the \mmstar relation, respectively. $y_{\text{lim}}$ is the limiting black hole mass at the median redshift $z = 5$. $\sigma_{x}$ and $\sigma_{y}$ are the measurement uncertainties of $\Mstar$ and $\Mblack$, respectively. $\sigma_{xy}$ is the covariance between the measurement uncertainties of $\Mstar$ and $\Mblack$. $\theta = \arctan(m)$ is the angle between the \mmstar relation and the x-axis. $F(t)$ is the cumulative distribution function of a Gaussian orthogonal to the relation, with mean $0$ (i.e., no deviation from the relation) and variance $\Sigma^2 + \nu$. The key terms of Eq. \ref{ln_L} and their statistical significance are the following:
\begin{itemize}
    \item $\Delta_i$: the orthogonal displacement of a data point from the fit line.
    \item $\Sigma_i^2$: the orthogonal variance, calculated by projecting the uncertainty covariance matrix (representing the two-dimensional Gaussian uncertainty of a data point) down to the line.
    \item $\phi_i$: adds extra weights to sources at lower stellar masses.
    \item $\eta_i$: corrects the flux-limited bias of JWST for the $\Ha$ line.
\end{itemize}

Note that $\phi_i$ and $\eta_i$ are modifications to the likelihood function in \cite{Hogg_2010}; we include them to address better the physical contexts we examine. The flux-limited bias correction term $\eta_i$ is derived and discussed in Nguyen et al. 2023 (in prep.).  We only consider the likelihood function within the stellar mass range of our sample ($7.79 < x < 10.66$) and above the limiting black hole mass (see Sec. \ref{subsec:JWST_sensitivity}).

Using the likelihood function described above, the local \mmstar relation parameters from \cite{Reines_Volonteri_2015} as priors for $m$ and $b$, and a flat prior for $\nu$ ($0 < \nu < 1$), we use the software package \texttt{emcee} \citep{emcee} to run a Markov Chain Monte Carlo (MCMC) algorithm, as outlined in \cite{Jonathan_Weare_2010}. We parametrize the \mmstar relation as
\begin{equation}
    \log \left( \frac{\Mblack}{\Msun} \right) = b + m \log \left (\frac{\Mstar}{\Msun} \right) \, ,
\end{equation}
and sample the $(b,m,\nu)$ parameter space. We assessed convergence by considering the integrated autocorrelation time to quantify the effects of the sampling errors on the output, as suggested by \cite{Jonathan_Weare_2010}. The execution of the algorithm is halted once the autocorrelation time converges within $0.1\%$, which takes $\sim 10,000$ steps. The first $1\%$ of steps during the burn-in period are not considered to avoid bias from the initial position (the parameters of the local \mmstar relation). Then, the posterior distributions calculated by the MCMC algorithm provide the median and 1$\sigma$ uncertainty of the parameters, providing an estimated high-$z$ \mmstar relation inferred from JWST data. The uncertainty of the inferred high-$z$ relation is assessed by sampling it 1000 times, with parameters chosen randomly within the 1$\sigma$ three-dimensional contours of the fit parameters.

To conclude, we briefly explore the effect of the flux-limited bias correction term $\eta$ in Eq. \ref{ln_L}.
If an infinite sensitivity characterized JWST, the best-fit line would simply pass through the data points (assuming equal uncertainties). Instead, JWST is characterized by a finite sensitivity to detect $\Ha$ emission; hence the distribution of observable sources is skewed towards higher black hole masses, as we neglect sources below the limiting mass, which are unobservable. Consequently, the best-fit line shifts lower because the limiting sensitivity increases the probability of sources with overmassive black holes. $\eta$ ensures that the inferred best-fit line is not overestimated as we consider the sensitivity limit of JWST.

However, because the JWST limit in $\Ha$ is deep ($\gtrsim$1 dex lower in the mass scale, see Fig. \ref{fig:limits}) compared to most black hole observations in our sample, $\eta$ only shifts the inferred \mmstar relation down by $\sim 0.2$ dex. In this study, the influence of $\eta$ is statistically insignificant compared to other uncertainty factors, typically on the order of $\gtrsim 0.3$ dex (e.g., the mass measurements, the intrinsic scatter, and the SMF). We also note that if the mass measurement uncertainties are underestimated, or the correlation between the uncertainties of $\Mstar$ and $\Mblack$ is overestimated, then $\eta$ would overestimate the inferred relation. A more detailed statistical analysis of the implications of JWST flux-limited bias for current and future high-$z$ observations is included in Nguyen et al. 2023 (in prep.).

\subsection{Estimate of the Number of Sources Observable}
\label{subsec:estimate_nr}

The $z \sim 0$ observations show that the root-mean-square (rms) scatter around the local \mmstar relation is $\approx 0.55$ dex \citep{Reines_Volonteri_2015}. Thus, assuming the local relation, we model the conditional probability density of detecting galaxies with a certain central black hole mass given its stellar mass, i.e., $p(\Mblack | \Mstar)$, as a Gaussian distribution with a standard deviation of $0.55$ dex around the \mmstar relation. Once we derive the \mmstar relation inferred from the $z \sim 4-7$ observations by JWST (see Sec. \ref{subsec:likelihood}), we can use the best-fit intrinsic scatter as the standard deviation around the relation to compute the expected number of observations.

Given the limiting black hole mass $M_{\text{lim}}$ at a certain redshift (see Sec. \ref{subsec:JWST_sensitivity}), the probability of detecting a source with $\Mblack < \widetilde{\Mblack}$, assuming $\Mstar$, is:

\begin{equation}
    {\cal P}(\Mblack < \widetilde{\Mblack} | \Mstar) = \int_{M_{\text{lim}}}^{\widetilde{M}_\bullet} p(\Mblack | \Mstar) \, d\Mblack \, .
\end{equation}
Hence, the surface density of sources with $\Mblack < \widetilde{\Mblack}$, at redshift $z$, is:
\begin{equation}
    \Psi(\Mstar, z) = {\cal P}(\Mblack < \widetilde{\Mblack} | \Mstar) \times V(z) \times \phi(\Mstar, z) \, .
\end{equation}
Thus, the expected number of sources with $\Mblack < \widetilde{\Mblack}$ observable by JWST with the $\Ha$ line, at redshift $z$, is:
\begin{equation}
    \int_{M_{\star \text{min}}}^{M_{\star \text{max}}} \Psi(\Mstar, z) \, d\Mstar \, ,
\end{equation}
where $M_{\star \text{min}}$ and $M_{\star \text{max}}$ represent the minimum and maximum stellar mass, respectively, in our considered sample. We limit our analysis within this stellar mass range to avoid overestimation bias from the lower stellar mass end. While more galaxies are distributed at lower stellar masses, they are dimmer and less likely to be observed by JWST, so it is reasonable to keep our estimates within the stellar mass range observed so far.

\section{Results}
\label{sec:results}

\begin{figure*}%
    \centering
\includegraphics[angle=0,width=0.49\textwidth]{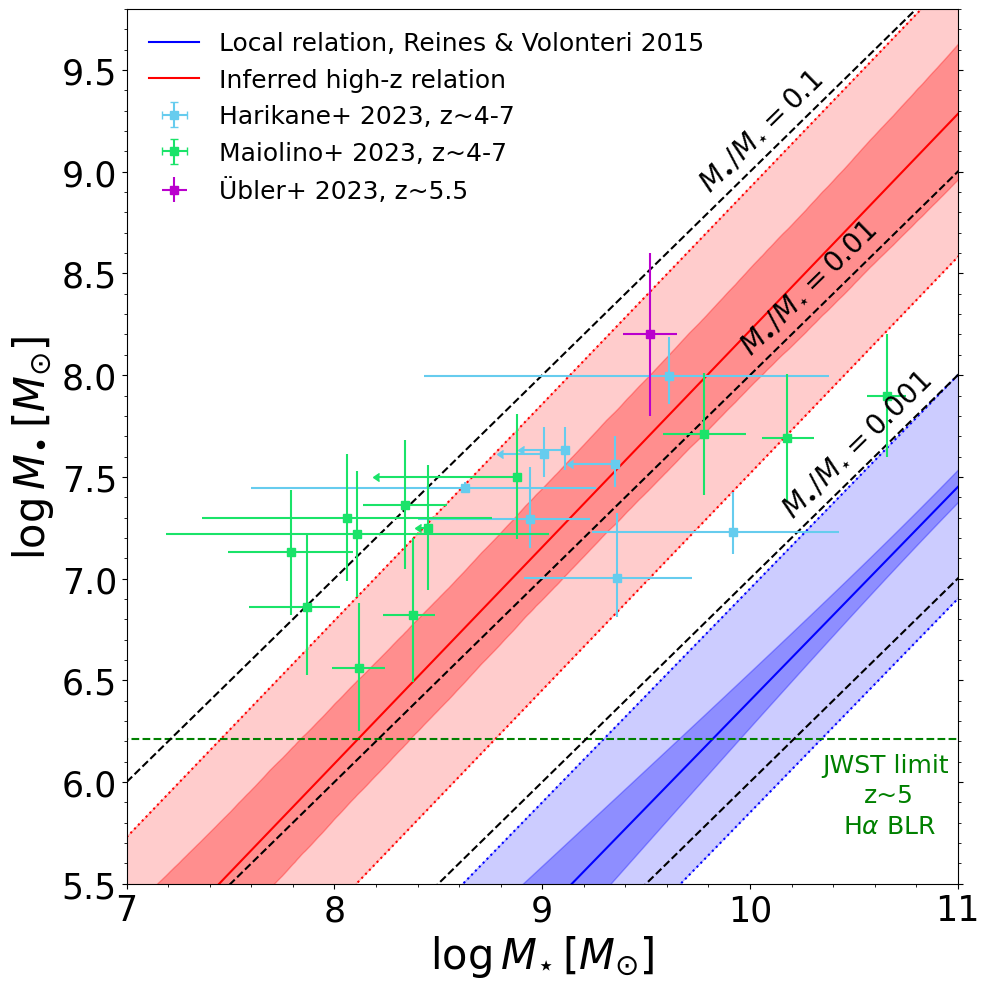} \hfill \includegraphics[angle=0,width=0.49\textwidth]{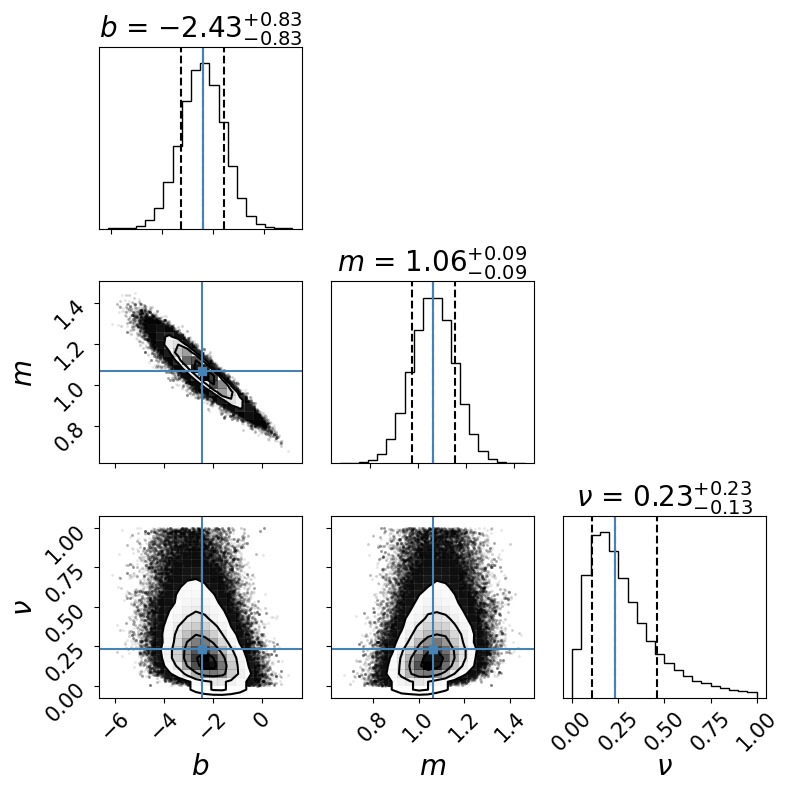}
    \caption{\textbf{Left panel:} in red, the high-$z$ \mmstar relation derived from JWST data at $z=4-7$. The dark-shaded 1$\sigma$ uncertainty region is derived from sampling $n=1000$ \mmstar relations randomly chosen within the 1$\sigma$ three-dimensional contour of the fit parameters. The red dotted lines and light-shaded region represent the best-fit intrinsic scatter of 0.69 dex (defined as $\sqrt{\nu} \sec \theta$, see Sec. \ref{subsec:likelihood}). \textbf{Right panel:} the posterior distribution of the parameters defining the \mmstar. The solid blue line and black dashed lines of the one-dimensional distributions denote the median and 1$\sigma$ uncertainties of the fit parameters. The solid contour lines in the two-dimensional distributions denote the 1$\sigma$, 2$\sigma$, and 3$\sigma$ joint uncertainties.}
    \label{fig:mmstar}%
\end{figure*}

In this Section, we use the framework detailed in Sec. \ref{sec:methods} to address two fundamental questions: (i) What can we infer with current JWST data regarding the \mmstar relation at $z=4-7$? (ii) What are the properties and observational perspectives of the population of black holes at $z=4-7$, especially in the low-mass range ($\Mblack \lesssim 10^{6.5} \Msun$)?

\subsection{The High-$z$ \mmstar Relation from JWST data}
\label{subsec:high-z_relation}

We use JWST data described in Sec. \ref{sec:data} and our theoretical framework detailed in Sec. \ref{sec:methods} to infer the \mmstar relation in the $z=4-7$ redshift range with an MCMC approach.

We seek a relation of the form $\log {\Mblack} = b + m \log \Mstar$, where both the black hole mass and the stellar mass are expressed in solar units. For our calculation, we consider the JWST sensitivity limit for the $\Ha$ line at $z = 5$ (median redshift of our data), the SMF, and the local \mmstar relation as a prior.

Figure \ref{fig:mmstar} shows our results. We find a high-$z$ \mmstar relation that is visually very different from the local relation by \citealt{Reines_Volonteri_2015}.
In particular, the high-$z$ relation indicates black holes + host systems where black holes are $\sim 10-100 \times$ overmassive with respect to stellar content of their low-$z$ counterparts. 

Our MCMC analysis yields the following values for the parameters: $b=-2.43^{+0.83}_{-0.83}$, $m = 1.06^{+0.09}_{-0.09}$, $\nu = 0.23^{+0.23}_{-0.13}$, where $\nu$ is the orthogonal variance of the best-fit intrinsic scatter. The standard scatter is $\sqrt{\nu} \sec \theta \approx 0.69$ dex, where $\theta$ is the angle between the \mmstar relation and the x-axis. Note that the standard scatter in the high-$z$ relation is larger than that of the local relation by \cite{Reines_Volonteri_2015}. The right panel of Fig. \ref{fig:mmstar} shows the posterior distributions.
The fact that the slope $m$ is consistent with unity indicates that the proportion between the black hole and stellar mass does not change with the size of the galaxy.
In summary, we derive, with high confidence, the following high-$z$ \mmstar relation from JWST data:
\begin{equation}
\begin{split}
    & \log \left( \frac{\Mblack}{\Msun} \right) = -2.43^{+0.83}_{-0.83} + 1.06^{+0.09}_{-0.09} \log \left (\frac{\Mstar}{\Msun} \right) \, \\
    & {\rm scatter \colon} \ 0.69 \, {\rm dex} \, .
\end{split}
\end{equation}

To test whether the local and the MCMC-inferred high-$z$ relations are statistically different, we perform a Welch's t-test. The Welch's t-test, contrary to the Student's t-test, does not assume that the two statistical populations have the same variance. For each value of the stellar mass $M_\star$, we ask whether the two values of the black hole mass $\Mblack$ predicted by the local and the high-$z$ relations are statistically compatible, considering their uncertainties expressed in terms of (corrected) standard deviations. To perform this test, we use the statistics:
\begin{equation}
    t(M_\star) = \frac{\Mblack^{\rm z}(M_\star) - \Mblack^{\rm loc}(M_\star)}{s_\Delta} \, ,
\end{equation}
where $\Mblack^{\rm z}$ is the black hole mass inferred in the high-$z$ relation, $\Mblack^{\rm loc}$ is the black hole mass for the local relation. Furthermore, $s_\Delta = \sqrt{s_z^2/n_z + s_L^2/n_L }$, where $s_z$ and $s_L$ are the corrected standard deviations computed for the high-$z$ and the local relation, respectively. The error sampling for the local and the high-$z$ relations are obtained with sample numbers of $n_L=244$ and $n_z=21$, respectively. 

Backed by this test, we conclude that the \textit{high-$z$ \mmstar relation is statistically different, at $>3 \sigma$, from the local one.}
Such a substantial deviation from the local \mmstar relation is not due to the JWST sensitivity limit. In fact, the MCMC relations inferred by taking or not taking the JWST sensitivity limit into account are only shifted by $\sim 0.2$ dex vertically from each other.

Several data points lay outside the standard scatter, which is expected. The scatter characterizes the intrinsic astrophysical processes that cause the deviation from the relation. These deviations cannot be properly accounted for by inference-derived uncertainties (see also the equivalent plot in \citealt{Reines_Volonteri_2015}).

Although the black holes in our sample span a mass range $ 6.5 < \log \Mblack < 8.2$, our inferred high-$z$ relation is visually steeper than the general distribution of data points. A combination of two arguments can explain this fact. First, the algorithm we used assigns more statistical weight to lower values of $M_\star$. Galaxies with a lower stellar mass are intrinsically more numerous and thus more representative of the general population. Hence, the larger $M_\star$ data points, which are lower than the fit, have less influence in flattening the inferred relation. Second, the data points at smaller $M_\star$ are closer to the JWST mass limit; hence, the algorithm improves their likelihood of hosting overmassive black holes. In principle, without a sensitivity limit, we might observe lower-mass black holes; this effect makes the slope of the inferred relation steeper than the distribution of data.

The offset between the local and the high-$z$ relations may partly be due to an underestimation of the stellar masses of the hosts \citep{Maiolino_2023_new}. A significant underestimation, sufficient to bring the systems back on the local relation, is, nonetheless, excluded for the following two reasons: (i) Even considering all the continuum emission in these systems to be produced by stars, they are still located many sigmas above the local \mmstar relation \citep{Maiolino_2023_new}; (ii) Significant deviations are not seen if these systems are compared with other local relations, such as the $\Mblack-\sigma$ and the $\Mblack-M_{\rm dyn}$ (see Sec. \ref{sec:intro}): observational biases should not play a significant role.

Hence, we conclude that the departure from the local \mmstar relation in the high-$z$ Universe is real, i.e., not due to a selection effect in flux-limited surveys. In other words, the high-$z$ galactic systems that JWST is detecting belong to a population of sources statistically different from the population from which the local \mmstar relation was derived. \textit{Black holes are significantly overmassive with respect to the stellar content of their host galaxies in the high-$z$ Universe.}

\subsection{Systematic Uncertainties}
\label{subsec:systematics}
We have shown that the empirically determined, high-$z$ \mmstar relation significantly differs from the local relation by $\sim 2$ orders of magnitude. Of course, this empirical determination is only as good as the underlying measurements of black hole mass, stellar mass, and their associated uncertainties.

Based on current data, we have no reason to believe that systematic uncertainties drive our findings. Typical uncertainties in the estimate of the black hole mass are of the order of $\sim 0.5$ dex \citep{Reines_2013, Reines_Volonteri_2015, Maiolino_2023_new}. Even if all the black hole masses were consistently overestimated by $0.5$ dex, the inferred relation would still be located above the local one in a statistically significant way. A typical source of systematic error in black hole mass estimates derives from the use of different $\Ha$ virial relations calibrated on different datasets. For instance, the mass estimated by \cite{Reines_Volonteri_2015} is $\sim 0.3$ dex above the relation by \cite{Greene_Ho_2004}. However, since we compare the high-$z$ data with the \cite{Reines_Volonteri_2015} relation, we eliminate this source of systematic error by recalculating all black hole masses using, consistently, that relation.

The black hole mass is estimated with the \cite{Reines_2013} calibration, which uses the width and luminosity of $\Ha$. The authors show a strong correlation between the luminosity of broad $\Ha$ components and the continuum luminosity of the AGN, allowing the use of the $\Ha$ luminosity to derive the radius of the BLR. While this method is less reliable than using the AGN continuum directly and is not yet calibrated for high-$z$ objects, it is the only black hole mass estimate available using current data.

Stellar masses derived from templates fitted to the spectra are more uncertain. For example, in the \cite{Maiolino_2023_new} sample, the most significant uncertainties associated with the stellar masses are $\sim 1$ dex, with most of them being significantly smaller. While some stellar masses may be underestimated, we argue that this cannot fully account for the $\sim 2$ orders of magnitude departure from the local relation. As commented by \cite{Maiolino_2023_new}, even the stellar masses computed without accounting for the AGN in the continuum modeling are still significantly above the local relation.

\subsection{Exploring the Low-mass Range of the Black Hole Distribution at $z = 4-7$ with JWST}
\label{subsec:low-mass}

JWST's strength resides not only in expanding the redshift horizon of black holes but also in peering below the tip of their mass distribution, discovering a potentially numerous population of lower-mass objects.
How low in black hole mass can we reach with JWST, given the inferred \mmstar relation at high redshift?

\begin{figure*}%
    \centering
\includegraphics[angle=0,width=0.49\textwidth]{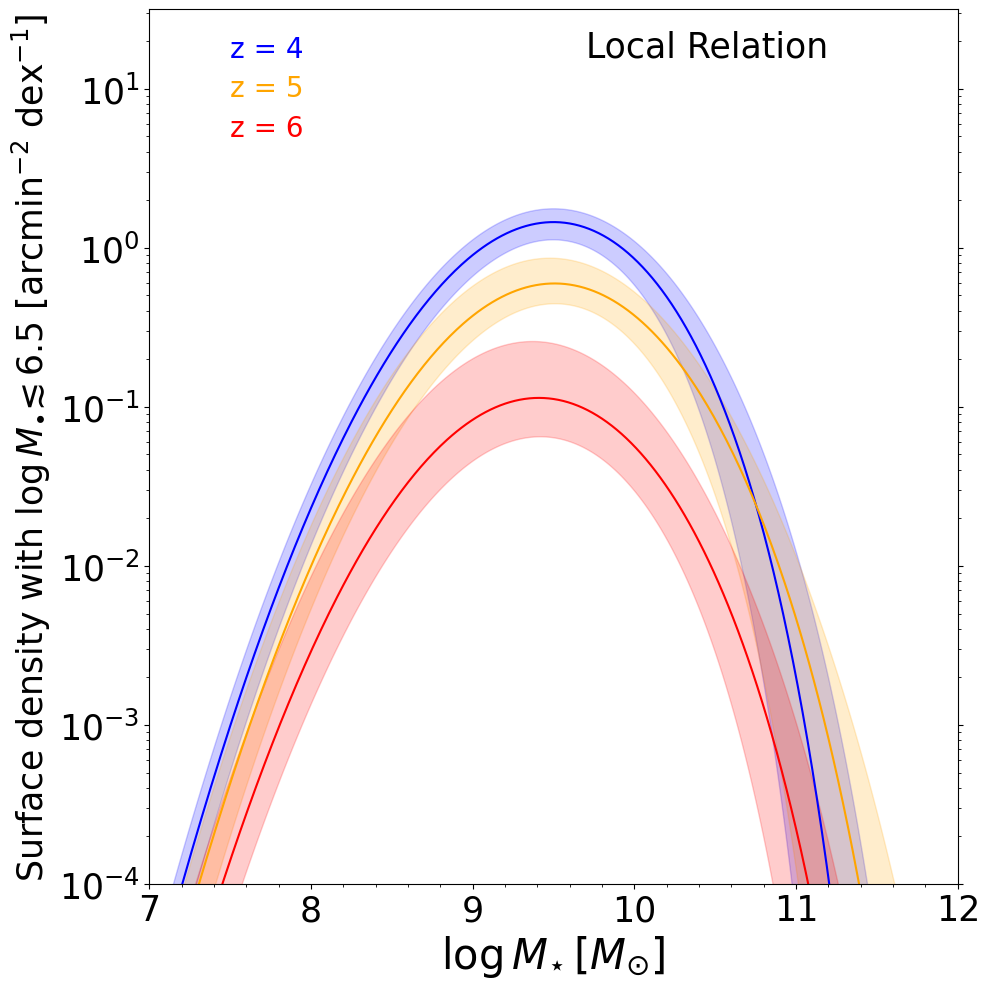} \hfill \includegraphics[angle=0,width=0.49\textwidth]{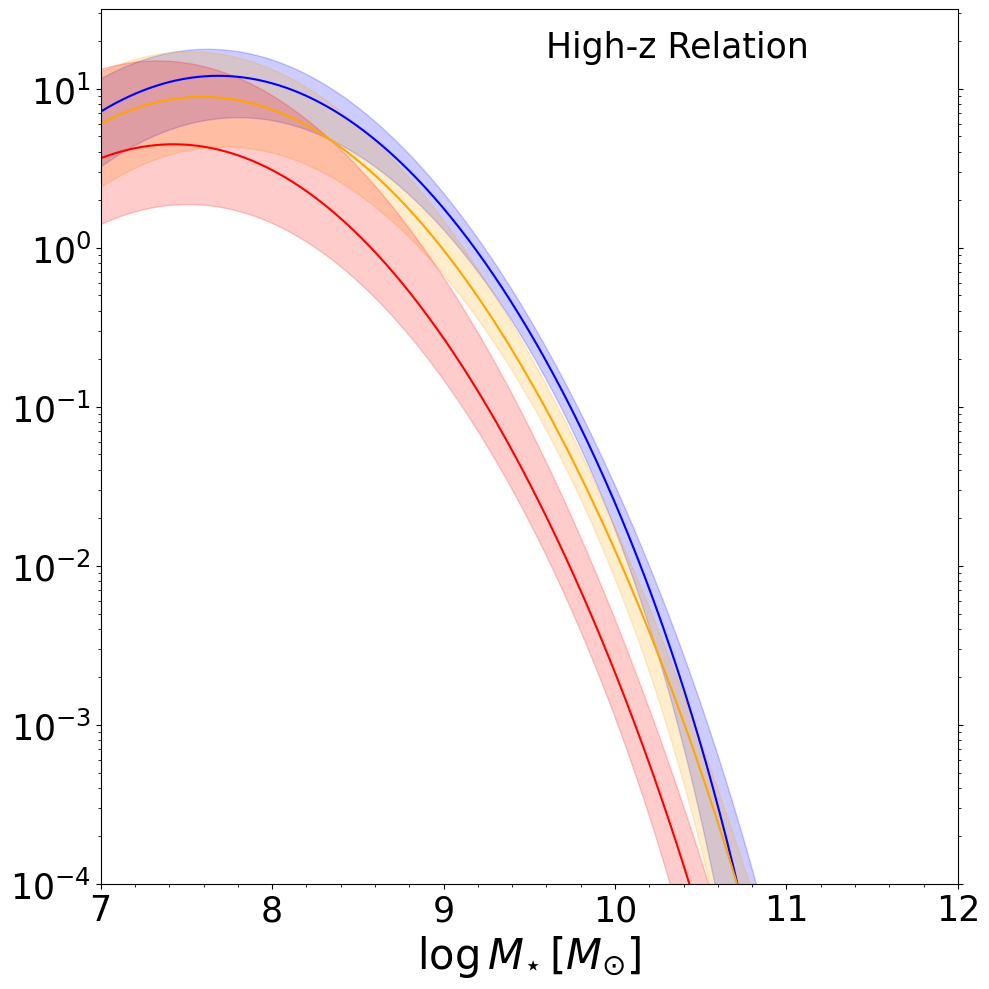}
    \caption{\textbf{Left panel:} expected surface density of sources with a mass $\Mblack \lesssim 10^{6.5} \Msun$, i.e., lighter than the smallest SMBH in our dataset: JADES 62309 \citep{Maiolino_2023_new}, assuming the local \mmstar relation. The shaded regions account for uncertainties in the black hole mass measurement using the $\Ha$ line (Eq. \ref{mass_Ha}) and the SMF (Eq. \ref{smf}). The expected surface density is provided for different redshifts $z \sim 4, 5, 6$, with a redshift binning of $\Delta z = 1$. \textbf{Right panel:} same as the left panel, but using our inferred high-$z$ relation. The peaks of the distributions are shifted to hosts with lower stellar masses, and the surface density of observable light-mass black holes is higher.}
    \label{fig:mass_lighter}%
\end{figure*}

In Fig. \ref{fig:mass_lighter}, we show the expected surface density of sources with a mass lighter than the smallest black hole detected so far in the redshift range considered, i.e., JADES 62309 with a mass of $\approx 10^{6.56} \Msun$ at $z=5.17$ \citep{Maiolino_2023_new}. Note that we exclude the lighter components of the candidate dual AGN in \citealt{Maiolino_2023_new}.
The calculation is performed at three different redshifts representative of our sample, assuming the local (left panel) and inferred high-$z$ (right panel) \mmstar relations.

Considering the local \mmstar relation, all three redshift distributions peak at a host stellar mass of $10^{9.5} \Msun$: it is more likely to detect small black holes in hosts of this stellar mass.
On the contrary, the inferred high-$z$ \mmstar relation predicts that the \textit{distribution of smaller SMBHs peaks at a host stellar mass of $\sim 10^{7.5}-10^{8} \Msun$} for redshifts $z = 4, 5, 6$. 

Since the high-$z$ relation indicates that the black holes are $\sim 10-100 \times$ overmassive compared to the stellar content of their low-$z$ counterparts (see Sec. \ref{subsec:high-z_relation}), black holes hosted by galaxies at lower stellar masses are massive enough to be detectable by JWST. In fact, recent JWST observations reported in \cite{Maiolino_2023_new} are distributed towards stellar masses of $\sim 10^{8} \Msun$. Our inferred high-$z$ relation suggests that JWST might be able to observe galaxies with even lighter stellar and black hole masses than sources detected so far. Expanding our observations towards the low-mass region of the \mmstar space will provide stronger constraints on the \mmstar relation and an experimental test of our inferred relation.

\begin{table}[]
\vspace{0.5cm}
\caption{Expected number of observable black holes with mass $\Mblack \lesssim 10^{6.5}$ (i.e., lighter than JADES 62309, the lightest black hole of the data set, see \citealt{Maiolino_2023_new}), in JADES and CEERS, for three redshift values. We assume the local \mmstar relation and the inferred high-$z$ relation. The redshift binning is $\Delta z = 1$.}
\centering
\hspace*{-2cm}
\begin{tabular}{lllll}
\hline
&  & \multicolumn{3}{c}{\textbf{Redshift}} \\ \hline
\textbf{Survey} & \textbf{Relation} & \multicolumn{1}{c}{\textbf{$z \sim 4$}} & \multicolumn{1}{c}{\textbf{$z \sim 5$}} & \multicolumn{1}{c}{\textbf{$z \sim 6$}} \\ \hline
\multirow{2}{*}{\textbf{JADES}} & Local & \multicolumn{1}{c}{$80^{+22}_{-23}$}  & \multicolumn{1}{c}{$34^{+16}_{-11}$}  & \multicolumn{1}{c}{$6^{+9}_{-3}$} \\
& High-z  & \multicolumn{1}{c}{$355^{+126}_{-133}$} & \multicolumn{1}{c}{$228^{+165}_{-97}$} & \multicolumn{1}{c}{$87^{+159}_{-45}$} \\ \hline
\multirow{2}{*}{\textbf{CEERS}} & Local & \multicolumn{1}{c}{$178^{+48}_{-50}$} & \multicolumn{1}{c}{$76^{+36}_{-24}$} & \multicolumn{1}{c}{$14^{+19}_{-7}$} \\
& High-z & \multicolumn{1}{c}{$887^{+316}_{-332}$} & \multicolumn{1}{c}{$570^{+412}_{-242}$} & \multicolumn{1}{c}{$217^{+396}_{-113}$} \\ \hline
\end{tabular}
\label{tab:surveys}
\end{table}

Table \ref{tab:surveys} reports the number of SMBHs with mass $\lesssim 10^{6.5} \Msun$ predicted in JADES ($\approx 45 \, \rm arcmin^{-2}$) and CEERS ($\approx 100 \, \rm arcmin^{-2}$) as a function of the three redshifts considered. 
Note that this calculation assumes an AGN fraction equal to unity, i.e., there is a 1-to-1 mapping between galaxies and black holes. As such, these numbers need to be considered as upper limits and include both Type 1 and Type 2 AGN.
We perform the calculation for the local \citep{Reines_Volonteri_2015} and the inferred high-$z$ \mmstar relations. 
We predict that current JWST surveys contain many more black holes lighter than $\sim 10^{6.5} \Msun$, especially at $z \sim 4$.
Remarkably, \cite{Maiolino_2023_new} reports a black hole with mass $\sim 4 \times 10^5 \Msun$, although this is part of a candidate dual AGN. This tentative detection implies that light black holes are within reach.

\textit{Our inferred high-$z$ \mmstar relation yields a higher value of detectable black holes with mass $\Mblack \lesssim 10^{6.5} \Msun$, compared to the local relation, at all redshifts}: we predict boosting factors of $5\times$ at $z=4$, $7\times$ at $z=5$, and $15\times$ at $z=6$. The difference is more evident at $z \sim 6$: a search of low-mass black holes in existing JWST surveys will convey a robust test of our \mmstar relation.

\subsection{General Mass Distribution of Black Holes}

\begin{figure*}%
    \centering
\includegraphics[angle=0,width=0.49\textwidth]{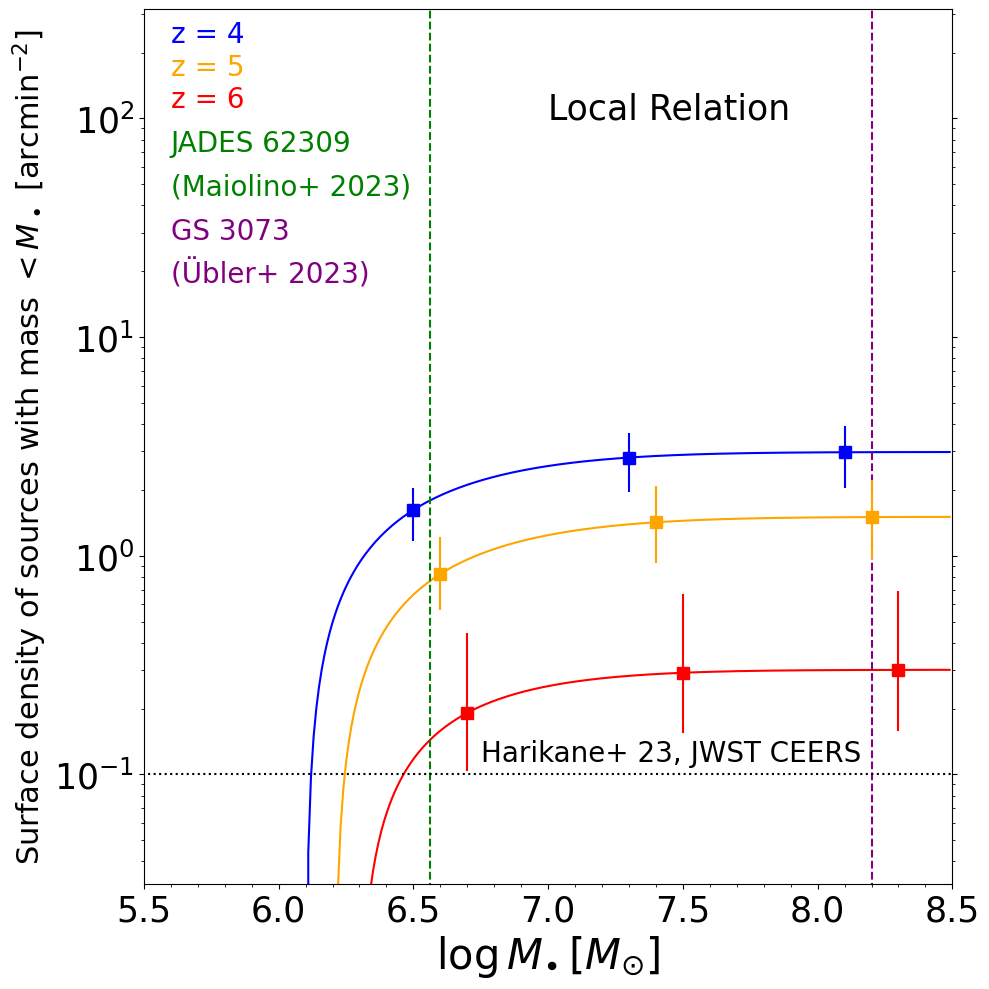} \hfill \includegraphics[angle=0,width=0.49\textwidth]{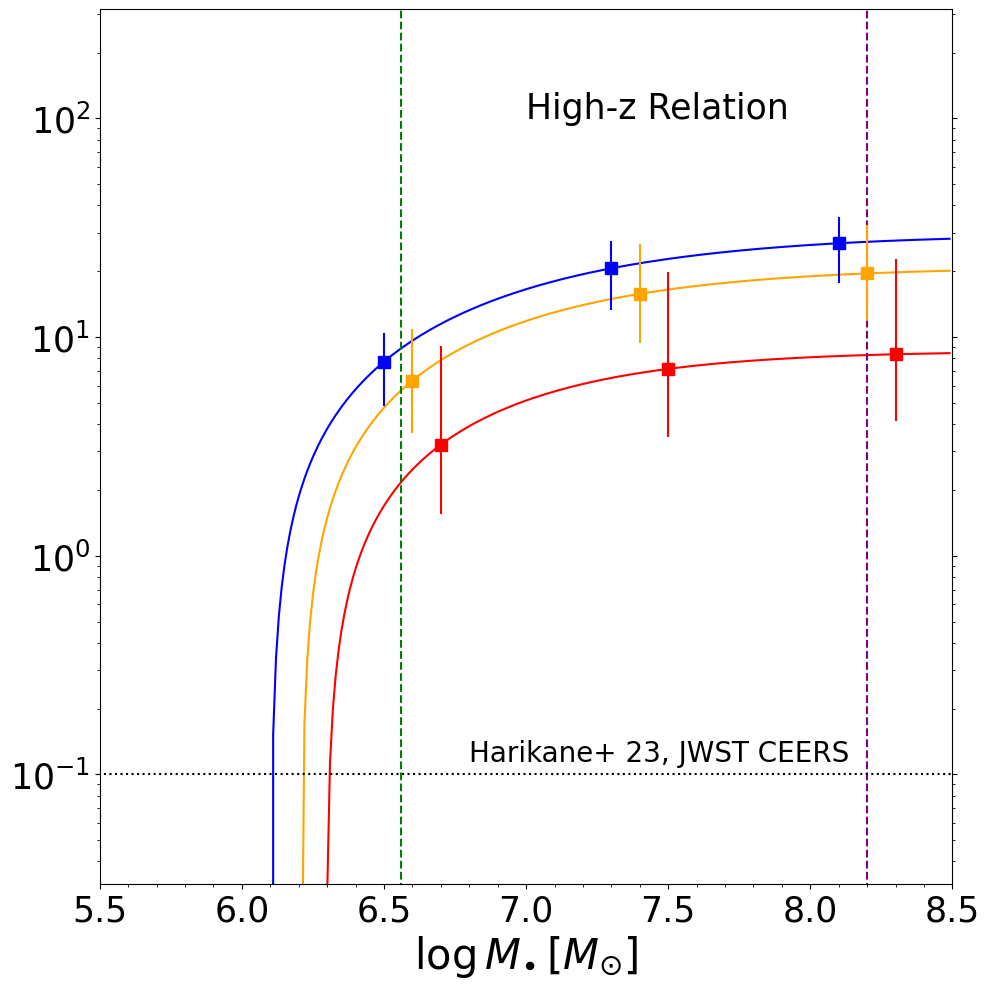}
    \caption{\textbf{Left panel:} expected surface density of sources with a mass $<\Mblack$, for $z=4,5,6$, assuming the local \mmstar relation. The two vertical lines indicate the lightest (JADES 62309, \citealt{Maiolino_2023_new}) and heaviest (GS 3073, \citealt{Ubler_2023}) black holes included in the data set. The horizontal line indicates the surface density of sources identified by \cite{Harikane_2023} in CEERS ($8$ sources in $\approx 72 \, \rm arcmin^{-2}$). The 1$\sigma$ error bars account for uncertainties in the black hole mass measurement using the $\Ha$ line (Eq. \ref{mass_Ha}) and the SMF (Eq. \ref{smf}).  \textbf{Right panel:} same as the left panel, but using the inferred high-$z$ relation. Again, the surface density of observable black holes is higher for all redshifts.}
    \label{fig:mass_distribution}%
\end{figure*}

Instead of searching for light black holes ($\Mblack~<~10^{6.5} \Msun$), we now investigate the general distribution in mass for black holes detectable by JWST with $\Mblack < 10^{8.5} \Msun$. Figure \ref{fig:mass_distribution} displays the expected surface density of sources, assuming the local (left panel) and the inferred high-$z$ (right panel) \mmstar relations.  

First, we note that the current searches in JWST surveys have likely found only a fraction of the SMBHs that are detectable; to visualize this fact, we show with a horizontal, dashed line the surface density of black holes detected by \cite{Harikane_2023} in CEERS ($8$ sources in $\approx 72 \, \rm arcmin^{-2}$). This result is independent of the assumption regarding the \mmstar relation.

For the general distribution of black holes, i.e., objects with a mass $\Mblack< 10^{8.5} \Msun$, our inferred high-$z$ relation predicts a boosting factor of $10\times$ at $z=4$, $15\times$ at $z=5$, and $30\times$ at $z=6$, compared to the local relation.

The fact that current JWST surveys are predicted to contain $\sim 10-100\times$ more black holes than the ones currently detected is remarkable. Possible explanations include:
\begin{itemize}
    \item Accretion at a lower Eddington ratio, which makes these AGN not easily detectable. The inferred accretion duty cycle would be $1\%-10\%$ (see, e.g., \citealt{Maiolino_2023_new} where they estimate it as $10\%$).
    \item A host galaxy that overshines the AGN, which makes them more challenging to detect \citep{Schneider_2023}.
    \item Obscuration of the central AGN, which makes them hard to identify with standard narrow-line diagnostics. In fact, high-$z$ AGN diagnostics overlap with the local ones.
\end{itemize}

Our predictions can also be compared with results from semi-analytical models. For example, \cite{Trinca_2023} estimates the number of black holes detectable in JADES and CEERS within the mass range $10^4-10^8 \Msun$. In particular, that study predicts that the black hole mass peaks in the mass range $10^6-10^8 \Msun$, which agrees with our general mass distribution using both the local and the inferred high-$z$ relation. 
However, \cite{Trinca_2023} predicts that the galaxy stellar mass distribution of observable sources peaks in the mass range $10^8-10^{10}$, which is similar to what we predict assuming the local relation.

\section{Discussion and Conclusions}
\label{sec:conclusions}

We used active $21$ galaxies discovered at $z=4-7$ by JWST in the JADES, CEERS, and GA-NIFS fields, hosting central SMBHs with $\Ha$-determined masses, to investigate, with an MCMC approach, the high-$z$ \mmstar relation and the mass distribution of black holes, especially its low-mass end.

Our main findings are summarized as follows:
\begin{itemize}
    \item The inferred high-$z$ \mmstar relation deviates significantly ($>3\sigma$) from the local relation. This fact is not due to selection effects in a flux-limited survey. We showed that the low-$z$ and the high-$z$ galactic populations are statistically different.
    \item Black holes are overmassive, by a factor $\sim 10-100$, when compared to $z\sim 0$ systems of similar stellar mass.
    \item Our inferred high-$z$ relation predicts, at $4<z<6$, $5-15 \times$ more black holes of mass $\lesssim 10^{6.5} \Msun$, and $10-30 \times$ more with $\lesssim 10^{8.5} \Msun$, compared to the local relation. The low-mass black holes will be preferentially found in hosts of $\sim 10^{7.5}-10^8 \Msun$ in stellar mass.
    \item Searching for low-mass ($\lesssim 10^{6.5} \Msun$) black holes in existing JWST fields can be a robust test of our \mmstar relation.
\end{itemize}

\begin{figure*}%
    \centering
\includegraphics[angle=0,width=0.85\textwidth]{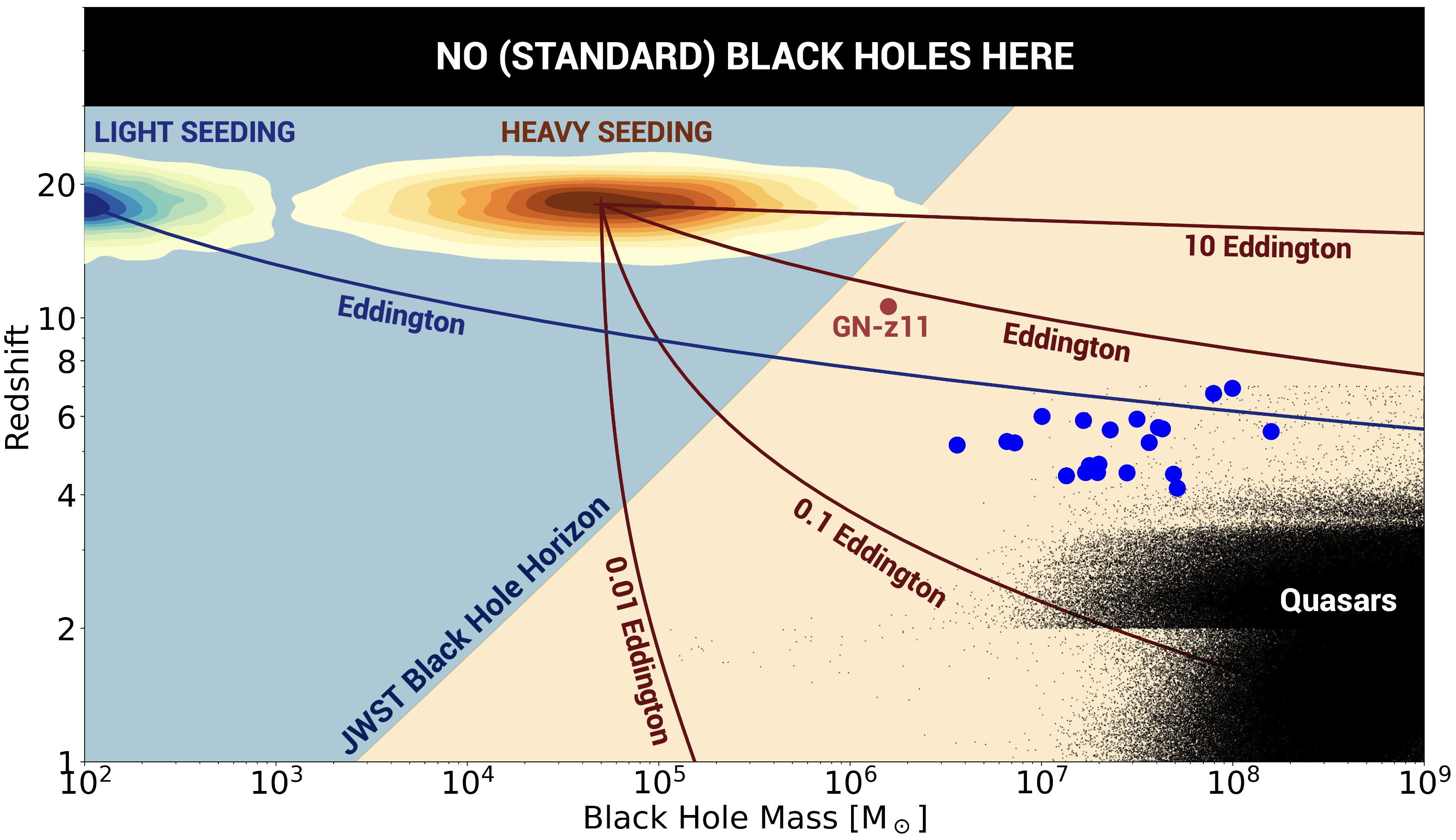}
    \caption{Qualitative representation of the current horizon within which black holes are detectable with JWST/NIRSpec. The galaxy + black hole systems used in this work are shown in blue. The farthest galaxy hosting a SMBH detected, GN-z11, is shown in red. The contours indicate typical heavy and light seeding distributions at $z > 10$. Some growth lines are also shown from the center of the seed distributions and assume several Eddington ratios (as indicated). The general distribution of lower-redshift quasars is displayed as a reference \citep{SDSS_2020, Fan_2022_review}.}
    \label{fig:horizon}%
\end{figure*}

Discoveries of single, high-$z$ sources containing black holes, such as GN-z11 \citep{Maiolino_2023}, are outstanding and prove the power of JWST in bringing the early Universe to light.
The uncovering of populations of high-$z$ systems hosting black holes is now starting to allow more thorough investigations into their general properties and to compare them with the reality of the local Universe.

In the local Universe, past the cosmic noon, galactic systems follow, with some scatter, all the standard relations: the \mmstar, the $\Mblack-\sigma$, and the $\Mblack-M_{\rm dyn}$. The high-$z$ Universe offers a different view. JWST is discovering galactic systems that follow the $\Mblack-\sigma$ and the $\Mblack-M_{\rm dyn}$ relations, but whose black holes are very overmassive with respect to the stellar content of their hosts, thus departing significantly from the \mmstar relation. Velocity dispersion and dynamical mass are, thus, better tracers of black hole growth because they directly depend on the central gravitational well. Black hole growth outpaces stellar growth at first. Then, possibly with the help of galactic mergers, the stellar mass catches up, and the systems end up on the \mmstar relation at low redshift. The only way to test this hypothesis is to further expand the black hole horizon.

Figure \ref{fig:horizon} offers a pictorial representation of the current black hole horizon, defined as the farthest redshift at which JWST can detect a black hole of a given mass, using the $\Ha$ broad emission line. For this qualitative estimate, we use a bolometric correction for the broad component of the $\Ha$ of $1/130$ \citep{Stern_Laor_2012, Maiolino_2023_new}.
The blue points represent the sources we used in this study, together with GN-z11 (in red), the farthest black hole with a spectroscopic confirmation \citep{Maiolino_2023}.

While the general population of quasars stacks up on the ``southeast'' of this chart (i.e., large black holes at lower redshift), JWST allows us to reach farther in redshift and lower in mass than ever. Because the SMBHs observed by JWST are less massive, they do not outshine their host galaxy. Hence, we can estimate most of these systems' stellar mass. 

Figure \ref{fig:horizon} also shows a qualitative representation of the mass-redshift distribution of light and heavy seeds (see, e.g., \citealt{Volonteri_2010, Ferrara_2014}).
GN-z11 can be interpreted as the evolution at a constant Eddington accretion rate of a typical heavy seed \citep{Maiolino_2023}.
Additionally, the current population of JWST sources analyzed in this work can result from the evolution, at a rate between $20\%$ Eddington and $80\%$ Eddington, of the population of heavy seeds formed at $z\sim 20-30$ \citep{BL_2000}. Alternatively, it can result from the evolution close to, or slightly above, Eddington of the population of light seeds.

Unfortunately, observations at even higher redshifts of massive black holes will be needed to pinpoint the mass distribution of the early seeds \citep{Pacucci_2022_search, Fragione_Pacucci_2023}. Alternatively, the shape of the high-$z$ \mmstar relation can inform us of the main formation channel of seeds \citep{Visbal_Haiman_2018, Pacucci_2018, Greene_2020_review, King_2021, Hu_2022, Koudmani_2022, Schneider_2023, Scoggins_2023_2}.
For example, recent simulations have shown that heavy seeds, with a typical mass of $10^5 \Msun$, can initially have a mass similar to or even larger than their hosts \citep{Scoggins_2023}. These systems remain overmassive, i.e., $\Mblack/M_\star \gg 10^{-4}$, until they merge with significantly more massive halos at $z \approx 8$. In the redshift range $z=4-7$ investigated in this study, \textit{we might be witnessing the slow but steady progress of these initially overmassive systems, formed from heavy seeds, towards the local \mmstar relation}.

We argued that the search in already existing JWST fields for low-mass black holes with $\Mblack \sim 10^6 \Msun$ provides crucial insights into the shape of the scaling relations at high-$z$.
Hence, JWST is not only expanding the horizon at which we can detect black holes but, in doing so, it is also enlarging our knowledge of the very early Universe.

\vspace{1.5cm}
\noindent \textit{Acknowledgments:}
We thank the referee for providing insightful comments on the paper. F.P. acknowledges fruitful discussions with Daniel Eisenstein and support from a Clay Fellowship administered by the Smithsonian Astrophysical Observatory. This work was also supported by the Black Hole Initiative at Harvard University, which is funded by grants from the John Templeton Foundation and the Gordon and Betty Moore Foundation. B.N. acknowledges support from an NSF Partnerships for International Research and Education (PIRE) grant OISE-1743747 and the Institute for Theory and Computation at the Harvard College Observatory. S.C. is funded by the European Union (ERC, WINGS, 101040227).

\noindent \textit{Software:} \texttt{Astropy} \citep{Astropy_2013, Astropy_2018, Astropy_2022}; \texttt{emcee} \citep{emcee}.

\bibliography{ms}
\bibliographystyle{aasjournal}



\end{document}